\documentclass[12pt,epsf]{article}
\usepackage{epsfig}
\usepackage{graphics}

\renewcommand{\thanks}[1]{\footnote{#1}} 

\newcommand{\be}{\begin{equation}}
\newcommand{\ee}{\end{equation}}
\newcommand{\bea}{\begin{eqnarray}}
\newcommand{\eea}{\end{eqnarray}}

\begin{document}

\begin{flushright}
SLAC-PUB-12505\\
May 2007\\
\end{flushright}

\bigskip\bigskip

\begin{center}
{\large\bf On Biology as an Emergent Science}
\end{center}

\bigskip\bigskip
\begin{center}
H. Pierre Noyes \footnote{\baselineskip=12pt Work supported in part
by Department of Energy contract DE--AC02--76SF00515.}\footnote{
\textbf{A preliminary version of this paper, entitled ``On Emergent
Science", was presented at the 28th Annual Meeting of the
Alternative Natural Philosophy Association, Wesley House, Jesus
Lane, Cambridge, England, August 2006. This version will appear in
the Proceedings of that meeting.}}\\
Stanford Linear Accelerator Center\\
Stanford University, Stanford, CA 94309\\

\end{center}

\bigskip\bigskip

\begin{center}
\bf{Abstract}
\end{center}

Biology is considered here as an ``emergent science" in the sense of
Anderson and of Laughlin and Pines. It is demonstrated that a
straightforward mathematical definition of ``biological system" is
useful in showing how biology differs in structure from the lower
levels in Anderson's ``More is Different" hierarchy. Using cells in
a chemostat as a paradigmatic exemplar of a biological system, it is
found that a coherent collection of metabolic pathways through a
single cell in the chemostat also satisfies the proposed definition
of a biological system. This provides a theoretical and mathematical
underpinning for Young's fundamental model of biological
organization and integration. Evidence for the therapeutic efficacy
of Young's method of analysis is provided by preliminary results of
clinical trials of a specific application of Young's model to the
treatment of cancer cachexia.

\newpage

\section{Why ``Emergent Science"?}

\subsection{Anderson, and Laughlin and Pines on ``emergence"}

In a famous paper Anderson\cite{A1972} has pointed out that there is
a natural hierarchy of scientific ideas. He starts with the usual
(reductionist) strategy of the search for the laws obeyed by the
elementary entities of physics, but then points out that the
possibility of reduction does \emph{not} imply constructivity.
Rather, if science $Y$ underlies some science $X$: ``The elementary
entities of science $X$ obey the laws of science $Y$". The $Y
\rightarrow X$ hierarchy Anderson proposes is: elementary particle
physics $\rightarrow$ solid state or many body physics; many body
physics $\rightarrow$ chemistry; chemistry $\rightarrow$ molecular
biology; molecular biology $\rightarrow$ cell biology; .... ;
physiology $\rightarrow$ psychology; psychology $\rightarrow$ social
sciences. Anderson goes on to state that ``... this hierarchy does
not imply that science X is `just applied Y'. At each stage entirely
new laws, concepts and generalizations are necessary, requiring
inspiration and creativity to just as great a degree as in the
previous one." I heartily agree!

Although my conventional scientific career (in elementary particle
physics) started out with the conventional scientific (reductionist)
assumption that the only way to solve a basic scientific problem was
to find the elementary entities, the laws they obey, and then
construct higher levels of science from that basis, I now realize
that I was mistaken. I have become convinced that 21$^{st}$ century
science will be most exciting and fruitful if its basic problem is
taken to be not only to find out if there are general hierarchy
bridging laws that connect each level to the next and lead to novel
types of complexity, but also if there are bridging laws which
overarch the ``elementary" bridging connection. This is one message
I read into the paper on emergent science by Laughlin and
Pines\cite{L&P2000} who explicitly start from Anderson's analysis.
Laughlin's book\cite{Laughlin2005} is criticized by
Leggett\cite{Leggett2005} under the title ``Emergence Is in the Eye
of the Beholder." However, I am still particularly impressed by the
fact that the values of $\hbar c/2e$ and $e^2/\hbar$ obtained by
electric measurements in complex systems can be obtained to much
higher accuracy than the values which can be obtained by direct
``elementary particle" measurements, despite the fact that the
details of the theories used to understand the complex systems
providing the data for these results have not reached consensus
agreement.

\subsection{The Organizing Principle of Darwinian Biology}

Laughlin and Pines\cite{L&P2000} also note that ``For the biologist,
evolution and emergence are part of daily life." As Fred Young
remarked when I  started discussing emergent science with him,
``Everything I ever said at ANPA (cf.\cite{YoungANPA}) was in the
direction of emergent science". This was sure to catch my attention
because for several years Fred Young\cite{Youngpc} has been trying
to explain to me how his thesis work\cite{Young77} is becoming more
and more important for him as an explanatory tool of use in
understanding how recent metabolic and physiological research all
fits together. When James Lindesay joined these discussions of
Fred's work, our joint understanding began to take shape as a
paper\cite{YLNL}. Briefly, I saw that Young's results and Lindesay's
mathematical deduction from them could be interpreted as the
starting point for adding the links ``cell biology $\leftrightarrow$
biological systems $\leftrightarrow$ ecological systems and
evolutionary biology" to Anderson's proposed hierarchy in a specific
way.

The careful reader will note that --- in contrast to the earlier
steps in Anderson's hierarchy --- I have used the symbol
``$\leftrightarrow$" for the connective between levels of the
hierarchy once biology enters the picture. The symbol
``$\rightarrow$" used by Anderson is an irreversible transition
which replaces the ``elementary particles"  of the lower level by
the laws they obey as the ``elementary entities" of the  new
phenomena which occur at the more complex (higher) level, and which
require the invention of new organizing principles, etc. which
``emerge" from the careful study of this richer world of ideas.
Biology arrived at its fundamental organizing principle by another
route. Some biologists did not even believe that the phenomena they
studied obeyed all of the laws of physics, in particular the second
law of thermodynamics! Further, it was found useful to ask what
\emph{purpose} a particular aspect of these complex biological
systems had ``evolved" to satisfy. This kind of \emph{teleological}
explanation had been banished from physics after a very hard
struggle, but its pragmatic usefulness in biology is hard to deny.
That a ``higher level" organizing principle can in fact lead to
\emph{deductive} and \emph{demonstrable} conclusions when applied
``top down" to a lower level \emph{biological} entity is one of the
points I wish to make below. This is why I replace ``$\rightarrow$"
by ``$\leftrightarrow$" in the hierarchy once we enter the
biological realm. Of course I must avoid the traps that lead to
error when teleological reasoning is used carelessly. I hope the
reader will reserve judgment as to whether I succeed in doing this
until my methodology is presented clearly. In particular I believe
that my methodology \emph{also} avoids the traps pointed out by
Anderson and by Laughlin and Pines, with whose basic conclusions I
do agree.

Biology has sometimes been called a ``Baconian" science in the sense
that it started by amassing all kinds of details and facts assumed
to be relevant to the subject and then induced general rules
governing these facts. This methodology is to be contrasted with the
tradition in the ``mathematical" sciences which started from the
astronomical practice of using numerical and geometrical models to
make testable predictions. Skipping over vital historical details,
this had the historical result that the ``physical sciences" came to
rely primarily on reductionism and hypothetical-deductive
methodology for testing. Chemistry started out as a Baconian
science, but began making the transition to a physical science in
the nineteenth century thanks to electro-chemistry, thermodynamics
and statistical mechanics. Quantum mechanics more or less allowed
that transition to be completed; this transition has often been used
as the leading example of the triumph of the
reductive-hypothetical-deductive methodology.

Biology has not as yet made much \emph{fundamental} use of
mathematics. Its greatest nineteenth century success was the
explanation of evolution via Malthus' observation that (in a stable
environment and over a sufficient period of time) a persistent
population \textbf{must} have birthrate = deathrate. Starting from
that deduction and with observations (in particular Darwin's) of
descent with modification, Darwin and independently Wallace came to
the conclusion that ``evolution by natural selection" is inevitable.
This was the \emph{non-quantitative} starting point for a scientific
``evolutionary biology". Since then this has remained unshaken as
the organizing principle of \emph{scientific} biology. Clearly ---
if my description of the history is roughly correct --- this is a
very different route to a basic organizing principle than the routes
followed in those ``physical sciences" which now rest on
hypothetical-deductive mathematical foundations.

One qualitative fact about biology which makes a methodological
difference between it and the physical sciences is that ``natural
selection" inevitably presupposes the existence of some sort of
\emph{environment} within which the biological systems evolve,
making it logically \emph{impossible} to discuss biological systems
without considering their interaction with that environment. I make
explicit use of this fact in my definition of what is meant by a
``biological system" in the sub-section which follows. A corollary
of this point of view is that the paradigm of most importance in
getting the study of biology off the ground is a persistent, evolved
system. We will see that this provides a \emph{reference state},
allowing fluctuations away from that reference state to be studied
\emph{quantitatively}.

\subsection{A Proposed Mathematical Definition of a Biological
system}

I \textbf{define} a persistent \emph{biological system} \textbf{B}
as a finite, countable \emph{population} of individual constituents
$C^B$ which in a suitable \emph{environment} at constant temperature
and pressure is a throughput (of molecules), steady state system
satisfying the first and second laws of thermodynamics. The
environment must supply food and fuel (\emph{F}) at a rate
sufficient to maintain the steady state. \textbf{B} acts
catalytically to convert the food and fuel into product molecules
\emph{P} which are retained by the individual \emph{living}
constituents and waste molecules \emph{W} which are disposed of by
the environment. The environment must also remove the waste heat
required by the second law in such a way as to maintain the
postulated constant temperature and pressure. The environment must
remove that selection of living individual constituents whose
disposal will maintain within the system (on average) a
\emph{constant distribution} of living constituents over all the
states which can occur during the \emph{lifetime} of any of them.
The environment must absorb all dead constituents. This implies that
``dead constituents" become part of the environment ``at death"and
are no longer part of \textbf{B}. In our context the (average)
number of (living) constituents satisfy the growth rate equation \be
\dot C^B = k_B C^B \ee and are said to be in a state of \emph{stable
population} (SP).

Note that $C^B$ is a ``counting number". As such, it is necessarily
\emph{dimensionless} in terms of a physicist's dimensional units of
mass, length and time ``$M,L,T$".  Then $\dot C^B$ and $k_B$ each
have the dimension of inverse time ($T^{-1}$). By taking Eq. 1 as
our defining equation for biology (in the context of a SP reference
state), we emphasize in a different way the importance of the
\emph{environment} in our definition of biology. Lacking any
evidence for a persistent \emph{physical} environment, any
biological \emph{system} satisfying our definition --- let alone its
individual constituents ---  \emph{must} have a finite lifetime.

\section{The Young Model for the Organization and Integration of Biological Systems}

\subsection{The Chemostat as  Paradigm for a Biological System}

Our definition of product molecules $P$ given in Sec. 1.3 allows us
to specify the distribution of living constituents by the (average)
number of product molecules they contain at any stage during the
life cycle of each constituent. We illustrate how this can be done
by narrowing the specific paradigm for a biological system used here
to a group of cells in a chemostat maintained in a SP state. For the
purposes of our theoretical analysis we assume that we can treat
each cell in the chemostat as a coherent combination of its chemical
constituents. Then we can use the symbol ``$C^B$" to stand for a
\emph{molecule} in the chemist's sense\footnote{A chemist's
``molecule" is a coherent structure which contains \emph{one or
more} chemical ``atoms", while a physicist usually thinks of an
``atom" as composed of still more elementary constituents, and of a
``molecule" as composed of \emph{two or more} atoms.}. This allows
us to write chemical equations (conserving the numbers of each type
of atom and the sum of their masses) connecting individual molecules
to cells, which we take to be one of the (implicit) axioms of
\emph{biochemistry}.

We are now dealing with a population of \emph{growing} cells inside
the chemostat absorbing nutrient molecules $F$ and producing product
molecules $P$ which are retained by the cell and waste molecules $W$
which are excreted into the solution surrounding the cell. Since the
cell eventually divides into two cells which --- at our level of
analysis --- are indistinguishable, we index the growing cells by
the number of product molecules $n_P$ they have added in the range
$N_P\leq n_P \leq 2N_P-1$. Cell division is then the irreversible
process

\begin{equation}
 C^B(2N_P) \Rightarrow 2C^B(N_P)
\end{equation}

\noindent The basic biochemical process in this context is

\begin{equation}
 F +C^B(N_P+n_P) \Rightarrow C^B(N_P+n_P+1) +W
\end{equation}

\noindent Consequently the chemical equation describing the
operation of the chemostat in this simplest case is

\begin{equation}
 N_F F +\Sigma_{n_P=0}^{2N_P-1}C ^B(N_P+n_P) \Rightarrow
 \Sigma_{n_P=1}^{2N_P-1}C ^B(N_P+n_P) + 2C^B(N_P) +N_W W
\end{equation}

\noindent which, by defining a \emph{complete population} of cells
(i.e. a population which, in the appropriate environmental context,
when supplied with $N_F$ nutrient molecules, can produce two clones
by cell division) as $\Sigma C^B \equiv \Sigma_{n_P=0}^{2N_P-1}C
^B(N_P+n_P)$, we can rewrite as

\begin{equation}
 N_F F +\Sigma C ^B \Rightarrow
 \Sigma C ^B + C^B(N_P) +N_W W
\end{equation}

\noindent or as

\begin{equation}
 N_F F {\Rightarrow \atop  \Sigma C^B} C^B(N_P) +N_W W
\end{equation}

\noindent A growing cell has to add $N_P$ product molecules to its
structure before it can divide and start the process over again. One
of those two copies (clones) must be removed at some subsequent time
(in its life cycle or when it dies); this pruning is required to
maintain the SP state. In this particulate description of the
overall process, any \emph{growing} cell will have to add each
individual product molecule \emph{sequentially}. We assume that the
context in which the equations apply is a \emph{through-put steady
state} (SP state). Then the rate at which the molecules of $F$ move
into the growing cell, the rate at which the molecules of $P$ join
the growing cell, the rate at which the cell divides into two clones
(beginning cells), and the rate at which one of these two growing
cells is eventually pruned from the cell colony are all the same.
That is

\be [\dot C^B]=k_B[C^B]; \ [\dot F]=k_B[F]; \ [\dot P]=k_B[P]; \
[\dot W]=k_B[W]\ee

\noindent Here the symbol $[X]$ ($X \in C^B, F, P, W)$ means the
\emph{concentration} (i.e mass per unit volume) of the substance
$X$, For small molecules (i.e. molecules whose atomic content and
(if needed) molecular structure are known) this mass is most
conveniently measured in terms of \emph{moles} (i.e. \emph{gram
molecular weights}). These equations immediately suggest that it may
be possible to treat concentrations of small molecules as biological
systems in an appropriate context. We develop this idea in the next
sub-section, in which we give precision to the concept of
\emph{metabolic pathway}.

The careful reader will have noted that we have use the symbol
$\Rightarrow$ denoting the \emph{irreversibility} of the chemical
reaction not only for cell-division (Eq. 2) but also for the
individual step (Eq. 3) in which the cellular environment
\emph{catalyzes} the transition from food molecule(s) to the product
and waste molecules. We assume that this can only happen when the
initial and final molecules are in the correct \emph{stoichiometric
ratios} (see next section). This is because we are interested in
this paper only in the passage of molecules through the cell (or to
their location within the cell) when this path does go through some
catalytic site (which we will call an \emph{enzyme}) that guarantees
that we are talking about a throughput steady state which is
\emph{far from equilibrium} and \textbf{not} about the equilibrium
states with which much of physical chemistry is concerned. Thus
there are no two-way transitions at the basic level and the usual
use of detailed balance rate constants is, from the start,
inapplicable. This brings us to the discussion of (enzymatic)
metabolic pathways in the next section.

\subsection{A Coherent Collection of Metabolic Pathways as a Paradigm
for a Biological System}

The food/fuel molecule or molecules $F$ that initiate the basic
process (Eq. 3) could have entered the cell at many different
places, and the waste molecule or molecules that complete the
process can leave the cell at many different places, but (in our
abstract model) the critical transition occurs at only one place
along the path(s) connecting the input and output surface patches,
namely where some enzyme $E_{F\Rightarrow PW}$ catalyzes the
reaction $h_F F\Rightarrow h_P P+ h_W W$. We call this ``one
dimensional" route through the cell a \emph{metabolic pathway} and
represent its action by the biochemical equation

\[
                      E\cdot h_F F
\]
\[
            \nearrow \ \ \ \ \ \ \searrow
\]
\be h_F F + E  \ \ \ \ \ \ \ \ \ \ \ \ \ \ \ \ \ \ \ \ \ \ \ \ E +
h_P P + h_W W \ee
\[
            \nwarrow \ \ \ \ \ \ \swarrow
\]
\[
                      E
\]

Eq. 8 represents the irreversible, catalytic action of a
\emph{single} enzyme molecule, which may dynamically change its
shape during the process but automatically resumes its initial shape
after the process is completed\footnote{This restoration of the
initial state of the enzyme provides one ``feedback" control
mechanism. \emph{Some} feedback control loop in the information flow
is \emph{required} for any persistent, self-organizing complex
system to exist.}. Note that, for the biochemical processes used in
our paradigm, this process \emph{must} occupy a (3+1)-dimensional
\emph{space-time volume} and hence \emph{must} be \emph{nonlocal}.
The numbers $h_F, h_P$ and $h_W$ \textbf{\emph{must}} be integers
because both the number of (chemical) atoms and the amount of
(chemical) mass are conserved in the process. Their ratios $h_{X/Y}
\equiv h_X/h_Y=(h_Y/h_X)^{-1}=h_{Y/X}^{-1}; X,Y \in F,P,W,...$ are
called \emph{stoichiometric ratios}. If we measure the
\emph{concentration} $[X]$ [which has \emph{physical} dimensions
$ML^{-3}$ (i.e mass per unit volume)] of any chemical substance $X$
in \emph{moles} (i.e. in gram-molecular weights per unit volume),
then the stoichiometric ratios are identical to the concentration
ratios. Then the equation also can be read as the number of moles of
each substance which will react in this way when catalyzed by one
mole of the enzyme. Note that we can \textbf{now} rigorously and
quantitatively bridge the \emph{small molecule $\leftrightarrow$
cell} mass magnitude gap by writing, as a corollary to Eq. 6

\be h_F N_F =h_P N_P + h_W N_W\ee

\noindent Note that this is an \emph{algebraic} equation connecting
positive definite \emph{integers} and \textbf{is not} a chemical
equation.

A few comments are needed here. Note that the $N_P$ apparently
independent metabolic pathways implied by Eq. 4 --- which are needed
in order to allow Eq. 7 to be treated as defining the hierarchical
nesting of a collection of biological systems --- must act
\emph{coherently}, at least at the conceptual level; this assumption
is also needed in order for the cell to be thought of as a coherent
chemical molecule. The conceptual advantage of this step is to allow
the very complicated process of cell growth and division to be made
into the simple doubling of the starting cell via the sequence of
steps (Eq. 4) that leads to cell division (Eq. 2). Then the rate
$k_B$ at which the transition occurs is a \emph{quantitative} and
experimentally measurable function of the concentrations of
\textbf{small} molecules of known structure called here $F$, $P$ and
$W$, even if we do not know the molecular weight of the enzyme
invoked by Eq. 8, or the details of how the catalytic result is
achieved, let alone knowing the molecular weight of the cell!

Some such critical conceptual step is needed in order for the
mathematical model we are constructing to be able to \emph{explain}
how chemostats can \emph{determine} empirically what function of
these concentrations the cell growth rate $k_B$ is. That such
functions are known is an empirical \textbf{fact}\cite{Mall...1966}.
It is this fact which allows us to go from it to a simple
mathematical formulation of Young's model. Explicitly we quote from
Fred's thesis (\cite{Young77}, p.1)

\begin{quote}
\ldots the value of \emph{$k_B$} is a reproducible function of the
medium composition\cite{Mall...1966} \ldots
\end{quote}

\noindent which we write formally as

\be k_B=K_B([F_1],[F_2],...,[F_{j}],...,[F_J]) \ee

\noindent Here the nutrients $F_{j}$ are distinguished from each
other by the unique enzymes $E_{j}$ which catalyze the
\emph{irreversible} reactions

\be h_{F_j} F_j \ \ {\Rightarrow \atop E_j} \ \  h_{P_j} P_j +
h_{W_j} W_j\ee

\noindent that remove $h_{F_j}$ molecules of $F_j$ from the
metabolic pathway and replaces them with product ($P_j$) and waste
($W_j$) molecules, conserving chemical mass and  atom flux. \emph{J}
is the number of \emph{types} of enzymes \emph{and} the number of
\emph{types} of metabolic pathways we consider important in any
particular analysis. $K_B$ is \emph{not} a function in the usual
mathematical sense. For us, if the values of the parameters are
known over the ranges of values and to the accuracy needed for our
immediate purposes, a ``table lookup" plus any well defined
``interpolation procedure" suffice to make this framework into a
\emph{testable theory} in Popper's sense.

Accepting that Eq. 10 is a reproducible \emph{empirical statement}
based on table lookup has important consequences. In that context
the inescapable \textbf{fact} is that all the numerical quantities
(in this case $k_B$ and each of the $[F_{j}]$) have an experimental
range of uncertainty. I formalize this fact by assuming that
\emph{whenever} we assert Eq. 10, we are claiming that there are
$2(J+1)$ numbers called $k_{min}, k_{max}, [F_j]_{min}, [F_j]_{max}$
such that for \textbf{any} choice of numerical values within these
ranges, no matter how correlated, the asserted equality provides an
acceptable representation of the data for our purposes. If this
statement becomes suspicious, the careful experimenter will look for
an explanation either in some source of systematic error, or some
theoretical constraint or possibility that has been ignored. Note
that in either case, these limits become testable hypotheses in
Popper's sense, and new experiments can either reduce the
experimental uncertainty or produce new empirical knowledge. This is
standard procedure in physics.

With this understood, we can use some hypothesis that makes nutrient
``$j$"  the ``most important" for the purposes of our analysis, and
formally ``invert" Eq.10 by defining

\be [F_{j}]=
(K_B)_j^{-1}(k_B;[F_1],[F_2],...,[F_{j-1}],[F_{j+1}],...,[F_J])\approx
(K_B)_j^{-1}(k_B) \ee

\noindent which means that, to the extent that the approximation is
valid, we can ignore what is going on in the concentrations of the
other nutrients and find some monotonically increasing function of
$k_B$, $k_{min}<k_B< k_{max}$ to fit the observed values of the
\emph{correlated} variation of $[F_j]$, for $[F_j]_{min} < [F_j] <
[F_j]_{max}$ (or \emph{visa versa}), --- i.e. $k_B =(K_B)_j([F_j])$.
With this basic phenomenology understood we can make testable
empirical hypotheses about and place reasonable theoretical
constraints on the concept of ``metabolic pathway" in the context of
a stable population of bacterial cells in a chemostat.

\subsection{Single Enzyme Control in a single metabolic pathway as
an irreversible transition}

The simplification of Eq. 8 for each emzyme/pathway \emph{j} to Eq.
11 allows us to compare it to the detailed model for catalytic
action in the irreversible reaction

\be 2 \ CO + O_2 \ \ {\Rightarrow \atop Cat.} \ \ 2 \ CO_2 \ee

\noindent as analyzed by Grinstein, et. al.\cite{Grinstein1989}. As
the authors note,

\begin{quote}
Since the reverse reaction $CO_2 \rightarrow CO + O$ is not allowed,
the system defined by the above rules cannot satisfy detailed
balance for any underlying Hamiltonian.
\end{quote}

\noindent which reinforces the remark already made at the end of
sub-section 2.1 that the processes we are considering \emph{cannot}
be described by the rules of equilibrium physical chemistry. It also
warns us (in our non-equilibrium context of irreversible, steady
state, throughput processes) that we cannot expect the essential
mathematics needed for theoretical biology to resemble the continuum
mathematics used in classical theoretical physics. I fear this fact
about biological systems is often ignored by biochemists analyzing
enzyme reactions \emph{in vitro}. The advantage Young has in basing
his model on chemostat data is that these empirical studies  are, in
fact, \emph{in vivo} experiments. They allow us to go
\emph{directly} from chemical measurements (concentrations of small
molecules) to a parameter ($k_B$) that measures the (average) time
it takes a \emph{living} organism to replicate itself in an
environmental \emph{context} that allows a biological system
composed of such organisms to achieve a persistent steady state
(SP-state).

The rules the authors\cite{Grinstein1989} refer to in the quote
given above describe the way to parameterize the rates at which the
incoming molecules attach to the catalytic surface, rearrange bonds
to form the product molecules of the outgoing gas, and the rates at
which the outgoing molecules detach. These details need not concern
us here, nor do the numerical methods Grinstein, et. al. are forced
to use because they lack a Hamiltonian model. What does concern us
is that the catalyzed transition $2CO +O_2 \Rightarrow 2CO_2$  is a
worked out example analogous to the way the $F$ molecules come along
the \emph{incoming} part of the metabolic pathway to a specific
enzyme and the $P$ and $W$ molecules leave on the distinctly
different \emph{outgoing} part of the same pathway. We \emph{could}
make a more detailed model of this process, but we \emph{are not
required} to do so in order to achieve our results. All we need
abstract from the complicated process that goes on in the
ill-defined space-time volume around the enzyme is the fact that
this transition separates the metabolic pathway into an incoming and
an outgoing part, and that it \emph{fixes} the stoichiometric ratios
of all the substances in this single metabolic pathway whose mass
flow is continuous through this volume.

The reason we are not concerned with the geometrical details is that
the basic equations (Eq. 7) are \emph{space-scale invariant}
\textbf{and} only depend on spacial averages (concentrations) as
functions of time. To smooth these out in the complicated region
where $F$ attaches to the enzyme, the enzyme rearranges $F$ into $P$
and $W$ and these leave, we assume this region has an average length
$L$. We assume that an average flow velocity for molecules along
this metabolic pathway through the cell can be defined by $v=k_BL$.
Then within this region, we can measure distance along the pathway
by a spatial coordinate $x= vt=Lk_Bt$ when $0 < x < L, 0 < t < T_B =
k_B^{-1}$. Assume that the pathway is in an SP-state (i.e.
$v=const.= k_BL = L/T_B$ ). \emph{Upstream} of this \emph{transition
region} (i.e $x<0$) the concentration $[F_j]$ must have a constant
\emph{input} value which we call $[F_j]_I$. \emph{Downstream} of the
transition region the concentrations $[P_j]$ and $[W_j]$ must have
constant \emph{output} values which we call $[P_j]_O$ and $[W_j]_O$.
We then know that the concentration $[F_j]$ must fall from its input
value $[F_j]_I$ to zero as it passes through the transition region.
This shows that we can \emph{always} describe the steady state
action of any enzyme which causes an irreversible phase transition
by

\be [\dot F_j] = - k_B [F_j] \ee

\noindent if we use the \emph{algebraic sign} conventions a) that
$k_B$ is a positive definite constant and b) that the time rate of
change of a concentration is \emph{positive} when it is the same as
the sign of the rate change of a \emph{growing} cell. Similarly
$[P_j]$ must start at zero at $x=0$ and rise to the output value
$[P_j]_O$ at $x=L$. Then a little thought tells us that if Eq. 8 is
used to represent an \emph{irreversible} transition, we must have
that

\be if \ [Y_j] \in [E_j], [E_j\cdot F_j], [P_j], [W_j], \ then \
[\dot Y_j] = k_B [Y_j] \ee

Now we must face up to the fact that,empirically, the chemostat data
often exhibit a substantial range of values of $k_B$, as is implied
by Eq. 10 and the quotation which it formalizes. Indeed, the
question of \emph{how} such a strictly correlated (by any set of
values or range of values for $k_B$ for which the quote and or Eq.
10 are a correct representation of the facts) can come about was the
problem Fred Young's thesis\cite{Young77} set out to solve.

Here the Darwinian organizing principle of natural selection comes
to our aid. Any organism in a biological system will benefit by
extending the range of the concentrations of nutrients which it can
tolerate and continue to reproduce, and the speed with which it can
make use of them in competition with other organisms or with
genetically modified members of it own species. But the metabolic
pathways \emph{within} each organism (having the same genome) will
gain collectively  (for its genotype) if their action is tuned to
make maximum use of the total supply which can be absorbed by the
organism as a whole. On both counts we expect an organism-wide
coordination to be selected for, and not just maximum range and
efficiency of the action of the individual catalytic pathways. As is
not surprising, from the point of view of the central dogma of
molecular biology, this coordination is provided by the genetic
control of the production of the enzymes themselves. Since this is
more easily explained by using the control mechanism discovered by
Fred Young in his thesis than by discussing a single enzyme pathway,
we now turn to that explanation in the next sub-section.

\subsection{Two Enzymes linked by a feedback loop in a single
metabolic pathway}

The basic feedback control loop for metabolic regulation which Fred
Young\cite{Young77} discovered, written as a chemical equation, is

\[
                      h_U U
\]
\[
            \nearrow \ \ \ \ \ \ \searrow
\]
\be h_I I + E_I \ \ \ \ \ \ \ \ \ \ \ \ \ \ \ \ \ \ \ \ \ \ \ \ E_O
+H_O O \ee
\[
            \nwarrow \ \ \ \ \ \ \swarrow
\]
\[
                      h_D D
\]

\noindent Note that this connects two irreversible enzyme-catalyzed
transformations, namely

\be h_I I + h_D D \ \ {\Rightarrow \atop E_I} \ \ h_U U \ \
{\Rightarrow \atop E_O} \ \ h_D D + h_O O \ee

\noindent Note that D --- which is analogous to the the enzyme E in
Eq. 8 (if we replace $E\cdot F$  by $D\cdot I =U$) --- is conserved
in the sense that it retains (in a SP-system) a constant
concentration. Note also that (in a SP-system) the net effect is to
transform I to O irreversibly, conserving mass and chemical atoms,
at a rate

\be k_B = {[\dot O] \over [O]} = - {[\dot I] \over [I]} = {[\dot U]
\over [U]} = {[\dot D] \over [D]} \ee

\noindent This is, of course, the same conclusion we reached about
the action of the individual enzymes (cf. Eq.'s 14 nd 15). This
means that we can go on connecting nodes (representing enzymes which
unequivocally direct mass flow in the direction defined by positive
growth rate or, in feedback links, unequivocally in the opposite
direction) in a way that will never upset the SP character of the
system \emph{provided} the environment is stable and we can prove
that the system is stable against ``normal" fluctuations in the
environment.

Further, Eq. 18 implies that ${[\dot O] \over [\dot I]} = -h_{O/I}$,
the negative of the stoichiometric ratio of the concentrations. Not
only is the rate of decrease of I precisely equal to the rate of
increase of O (a fact we could derive directly from chemical mass
and atom conservation), but if we think of the cell as a factory for
the production of O, ratios of the stoichiometric coefficients in
Eq. 16 could serve as the set-points for some rate control system
that optimizes the use of resources I to the rate at which they are
provided. This is obvious to a chemical engineer. That natural
selection has ``engineered" such a system is a deduction from the
Darwinian organizing principle.

Fred Young's approach is to ``reverse-engineer" the data on the
concentrations in steady state throughput experiments using what is
known about the structure and working of the cell so as to tease out
how the control system operates. One advantage of using his control
cycle, rather than concentrating on the genes (and hence the
enzymes) \emph{directly}, is that his control loop allows this to be
done using only the concentrations of the small molecules as the
empirical starting point. As he notes (\cite{Young77}, p.8): ``The
interrelationships between cellular components that define the
steady-state and illustrate the scope of regulation which is
independent of specific [genetic] induction-repression mechanisms
have been comprehensively tabulated."

One general mechanism for cell-wide rate control recognized by Young
is the relation between protein synthesis and ribosome synthesis
when both are thought of as a function of $k_B$ (cf. \cite{Young77},
Fig. 7, p.37 and related text). For a stable population of growing
cells, and a large range of values of $k_B$, the rate of protein
synthesis (per genome equivalent of DNA) is constant, whereas the
relative rate of ribosome synthesis is a rapidly increasing function
of $k_B$. Clearly the value of $k_B$ where these two curves cross is
a ``rate control set point" .

Why is this true? The proteins are manufactured by the ribosomes.
The particular protein called for is coded on an ``instruction tape"
(messenger ribonucleic acid --- mRNA). This tells the ribosome which
of the 20 possible amino acids to attach next onto the growing
protein (polypeptide) chain.  An ``expressed" DNA-gene uses
(ignoring ambiguities in the code) one of 20 ``three letter codons"
(corresponding to the 20 amino acids which can be used to make a
protein chain) to provide the information added sequentially to the
mRNA instruction tape. The fact that the concentrations of
ribosomes, ribosomal nucleic acid (rRNA), mRNA and tRNA are
\textbf{all} proportional to $k_B$ then tells us (accepting the one
gene -  one enzyme doctrine and still subtler
approximations\footnote{[DNA is a shorthand for
\emph{deoxyribonucleic acid}, The transfer RNA (tRNA) --- with 20
varieties --- transfers \emph{uniquely} the sequential information
from an expressed mRNA transcript of a DNA ``structural" gene one
codon at a time by having one end which attaches by complementary
base pairing to the RNA codon and picks up on the other end the
cognate amino acid which is added to the growing polypeptide chain.
If this apparatus worked perfectly there would be approximately one
``constant" rate (with a fine structure of 2 or 20 or some number
less than 64 rates) for this process. But the mRNA can itself get
degraded at some rate between the time when the information is
transferred to it physically and the time when it is read.
Consequently information transfer and the transfer of the material
coding of that information can have different average rates.
Fortunately, for the purposes of this paper, we can ignore these
complexities.}) that we can expect the concentration of any enzyme
(a ``large molecule" made up of one or more polypeptide (protein)
chains) produced by this machinery to also be proportional to $k_B$.
The fact that the (relative) rate of protein synthesis as a function
of $k_B$ is constant (in the range where it crosses the rate of
ribosome production) can be interpreted as due to the likelihood
that the rate of transcription for any of the codons that specifies
any amino acid is approximately the same as for any other codon.

The next step is to note that the amount of the enzyme synthesized
is controlled by the expression of the gene and that this, in turn,
is controlled by the operator-promoter region of the gene. These
controls can be either positive (enzyme induction) or negative
(enzyme repression) and can be effected by a change in the
concentration of any appropriate small molecule or protein in the
metabolic pathway upstream of the enzyme in question, even though it
is not directly involved in the $I \Rightarrow O$ catalysis. [Such a
change can even ``turn off" the gene completely and hence form the
starting point for a \emph{concentration threshold}-controlled
on-off ``switch". We will mention such switches in the discussion of
the cancer cachexia treatment, but not model them in this paper.]

Two applications of this control loop are discussed in
\cite{Young77}. The first is glucose metabolism. For it ``I" is some
phosphate in the food which is picked up by adenosine-diphosphate
(ADP) [identified with ``D"] to form adenosine-triphosphate (ATP)
[identified with ``U"], which then passes on the phosphate to some
downstream product [identified with ``O"] and returns ADP as the
feedback which completes the cycle. Thus there is an internal
``set-point" for the internal control loop --- namely the
stoichiometric ratio $h_{U/D}$ --- and an external set-point,
$h_{I/O}$. Both are under the genetic control of the two genes
associated with the two enzymes $E_I$ and $E_O$. The second
application is the addition of a monomer as the next link in the
chain of a growing polymer. The second example is illustrated in
\cite{Young77}, Fig. 5, p.28, which is reproduced below, together
with its figure caption as Figure 1. I trust that, after the above
sketch of how the whole thing works, this captioned figure is
self-explanatory.

What Fig. 1 does not point out is that the genetic control mechanism
acts on a long time scale, appropriate to a secular change in the
rate and/or amount at which food is flowing into the system from the
environment, while the internal control loop acts on a shorter time
scale which can smooth out short term fluctuations in the
concentrations due to other causes. That this feedback is
\emph{stable} follows immediately from the irreversibility of the
direction of flow at the two enzymes. These absorbing state phase
transitions (from $M + X$ to $(X\cdot M)$ and from $(X\cdot M)$ to
$X+ P$ --- with the return control loop path that takes $X$ ``back"
to the initiating phase transition in place) also enforce the
stoichiometric set points for both the interior and the exterior
rate control. The rate controlled \emph{range} of $k_B$ comes from:
a) the fact that the number of ribosomes per cell increases with
$k_B$; b) the fact that the range of $k_B$  is limited at the upper
end by the maximum number of ribosomes which the cell can hold in a
steady state; c) the fact that the range of $k_B$ is limited at the
lower end by the minimum amount of nutrient which will allow, at
least, the minimum stable number of cells to maintain themselves in
the chemostat at the flow rates for nutrient input and for the
solvent carrier input.

\newpage

\begin{figure}[ht]
\centering
\includegraphics{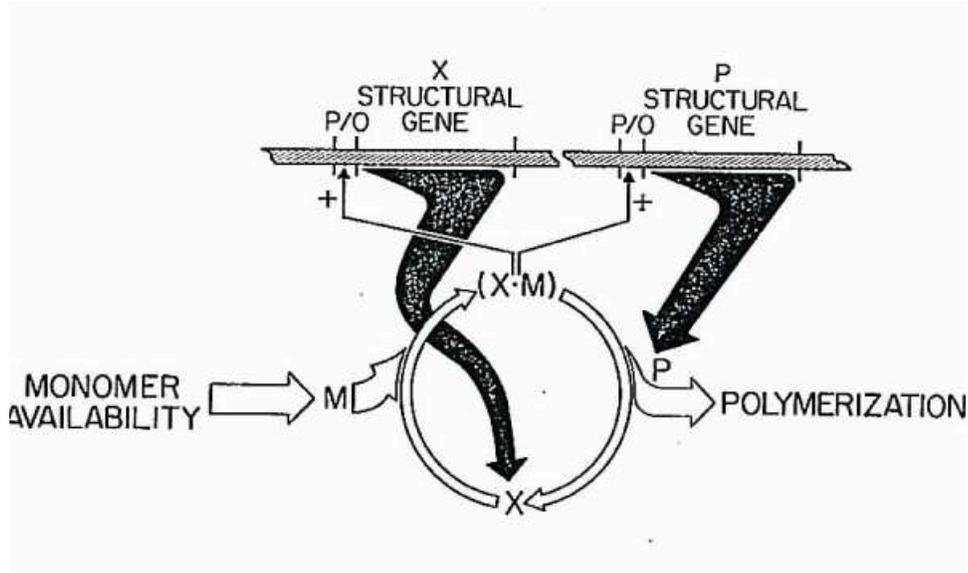}
\caption{\baselineskip=12pt Representation of how monomer charge can
be maintained at various steady-state growth rates through the use
of rate effectors.  When rates of monomer producing and utilizing
reactions are balanced the concentration of $(X\cdot M)$ is
proportional to the rate of synthesis of $(X\cdot M)$. $(X\cdot M)$
is a positive effector for synthesis of both $X$ and $P$. In this
way the rates of synthesis of $P$ and $X$ will be proportional to
the rate of synthesis of $(X\cdot M)$. During non-steady-state
conditions the concentration of $(X\cdot M)$ will no longer be
proportional to the rate of synthesis of $(X\cdot M)$ and
consequently rates of synthesis of $P$ and $X$ will no longer be
equal to the rate of monomer availability. P/O is the operator
promoter region of the structural gene. $X$ represents a carrier of
monomer units. (The thickness of the arrow is not meant to reflect
the relative reaction rates.)}
\end{figure}

The general applicability of Fred Young's general in-out, up-down
rate control loop feedback model, i.e.

\[
                      U
\]
\[
            \nearrow \ \ \ \ \ \ \searrow
\]
\be \rightarrow I \ \ \ \ \ \ \ \ \ \ \ \ \ \ \ \ \ \ \ \ \ \ \ \ O
\rightarrow \ee
\[
            \nwarrow \ \ \ \ \ \ \swarrow
\]
\[
                      D
\]

\noindent to most (possibly all) biological systems, and to many
well-modeled physical systems that provide significant analogies for
biological systems, may not be obvious. After all it was only first
discovered in \emph{E-coli} metabolism. But careful perusal of
Young's thesis\cite{Young77}, should begin to remove doubt on that
score. The thesis was deliberately undertaken, not to solve a
specific problem but to find a mechanism that could account in a
general way for how rate control of metabolism can lead to multiple
rates of growth in stable populations and for growing systems
exhibiting balanced exponential growth. Subsequent developments in
biology and other fields provide ample evidence for the fact that
Young's model is \emph{ubiquitous} in its applicability. This topic
will be discussed in\cite{YoungNat}, where the connected chain:
non-equilibrium steady state $\rightarrow$ absorbing state phase
transition $\rightarrow$ allometric scaling laws $\rightarrow$
fractal scaling $\rightarrow$ Kolmogorov scaling is developed. The
abstract of an earlier talk by Young on this subject at an
international meeting in Shanghai is presented here as Appendix 1.

Although the primary purpose of this paper was to prepare a
mathematical and logical basis for the Young model, whose technical
structure will be presented elsewhere\cite{YLNL}, we wish to also
take this occasion to point out that, once the biological principles
are understood, the top-down analysis of metabolic pathways which
Young's general model makes possible did not have to wait for
mathematical development in order to be applied. The underlying
logic and analytic framework have been used as the basis for
research to identify, select and assemble data from the published
literature. This data can then be used to create models for Cancer
Cachexia and other diseases. This systematic approach, starting from
Young's 1977 thesis\cite{Young77}, developed by Young and
collaborators and now called HiNET, then allows combined drug
therapies appropriate to these diseases to be constructed. The
technology has led to a venture capital backed company with a
portfolio of high potential ideas, one already in FDA-approved Phase
2 trial and two more likely to enter trial in 2008.

The problem of cancer cachexia can be briefly described as follows.
Normal nutrition for our species and many others has a
replete-hungry cycle with on-off switches changing the metabolic
pathways between the two stable states. In certain shock states,
there a great need for nutrition at any cost and the body in these
states starts eating anything inside it, including its own
structure. Normally this state turns off when the danger is past,
but cancer and some cancer therapies can produce a shock state that
does not return to normal; consequently the body wastes away even
though ample nutrition is provided by injection into the veins.
Using his analysis, Young found a way to treat the patient with
combinations of FDA-approved drugs. They alter the concentrations of
small molecules in the direction  which returns the body to normal
nutritional states and solves the problem. Preliminary clinical
trials were successful and second stage trials were approved. An
older short report of this is given in Appendix 2.

In conclusion, I believe that Young's control loop feedback in-out:
up-down cycle model for a throughput system is a good candidate to
become an emergent fundamental law of biological systems going
beyond the Darwinian organizing principle. Using such control loops
as coupled nodes in a hierarchical model for top-down analysis of
functional metabolic pathways, of which the first examples are
Young's HiNET models, bids fair to become a fruitful research tool
for uncovering novel emergent biological organizing principles
during the 21$^{st}$ century.

\section{Acknowledgments}

This paper rests primarily on the decades of work by Fredric S.
Young on his unique approach to the problem of the organization and
integration of biological systems. I am most indebted to him for his
invitation to participate in this research. I am also indebted to
James V. Lindesay for his collaboration on clarifying the
mathematical structure of the feedback system involved in the
two-enzyme control loop, and to both of them for the three-way
discussions we had during 2004-2006. I owe much to the the
clarification of the logical and philosophical structure of what we
are attempting to do provided by Walter R. Lamb while he was still
with us.

\newpage

\section{Appendix 1: The Universal Modular Organization of
Hierarchical Control Networks in Biology}

\begin{center}
Fredric S.Young\\
Vicus Biosciences [now Vicus Therapeutics, LLC]\\
\end{center}

Progress in the physical sciences have always involved conceptual
and theoretical simplification and unification. Modern biology has
resisted this tendency and has focused almost completely on the
details.  The sequencing of the human genome has not been translated
into comprehensive models and has not led to new therapies. Using
reverse engineering, we have abstracted a theoretical description of
the universal modular organization of biological control systems
which are modeled as the construction of a fractal representing the
hierarchical control network or HiNet. Disease therapy becomes a
problem of shifting the state of the HiNet to a configuration closer
to normal homeostasis. This has enabled the rational and systematic
developmet of combination therapies for clinical trials. An emphasis
on energy and control manifolds connects this approach to
catastrophe theory. The modularity of HiNet allows a hierarchical
network decomposition and modeling of local processes on low
dimensional control manifolds. Modeling of the integrated global
organization of a biological system requires control spaces of many
more dimensions than 3 as stated by Thom in \emph{Structural
Stability and Morphogenesis}. A HiNet model of allometric scaling
supports the recent application by Ji-Huan He of El Naschie's
$\varepsilon^\infty$ theory to biology. (Abstract of paper presented
at the 2005 International Symposium on Non-Linear Dynamics:
Celebration of M.S. El-Naschie's 60 Anniversary, December 20-21,
Shanghai, China)

\newpage

\section{Appendix 2: The Obsolescence of Reductionist Biology:
Systems Biology Modeling and Cancer Cachexia Therapy Development
Based on Emergent Patterns of Organization Rather Than on Genes and
Molecules}

\begin{center}
\emph{Dr. Fredric Young, Chief Scientist, Vicus Therapeutics, LLC}
\end{center}

Vicus has developed a hierarchical network (HiNET) model of emergent
patterns of organization based on principles of self-organized
criticality, phase-transitions, integral control and reaction
blocks. We will describe our HiNET model of cancer cachexia, a
catastrophic wasting disorder secondary to advanced cancer, and its
predicted EKG-based biomarkers and reaction-block drug targets. We
will show data from our retrospective and prospective VT-122
clinical trials and contrast our clinical results with previous
failed attempts targeting specific dysregulated pathways and
proteins. (Abstract of paper presented at \emph{Beyond the Genome
2006: Top Ten Opportunities in the Post-Genome Era}, June 19=21,
2006, Fairmont Hotel, San Francisco, California.)

\end{document}